\documentclass[prl,twocolumn,showpacs,superscriptaddress,amsmath,amssymb]
{revtex4}
\usepackage{bm}
\usepackage{graphicx}

\newcommand{\ket}[1]{|#1\rangle}
\newcommand{\bracket}[1]{\langle #1 \rangle}
\newcommand{\eps}{\varepsilon}
\newcommand{\intk}{\int_\text{BZ}\frac{d\bm k}{(2\pi)^d}}

\renewcommand{\cal}{\mathcal}

\begin{document}

\title{Theory of electric polarization induced by inhomogeneity in crystals}

\author{Di Xiao}
\affiliation{Department of Physics, The University of Texas at Austin, 
  Austin, Texas 78712, USA}

\author{Junren Shi}
\affiliation{Institute of Physics and ICQS, Chinese Academy of
  Sciences, Beijing 100080, China}

\author{Dennis P. Clougherty}
\affiliation{Department of Physics, University of Vermont, Burlington,
  Vermont 05405, USA}

\author{Qian Niu}
\affiliation{Department of Physics, The University of Texas at Austin, 
  Austin, Texas 78712, USA}

\date{\today}

\begin{abstract}
We develop a general theory of electric polarization induced by
inhomogeneity in crystals.  We show that contributions to polarization
can be classified in powers of the gradient of the order parameter.
The zeroth order contribution reduces to the well-known result
obtained by King-Smith and Vanderbilt for uniform systems.  The first
order contribution, when expressed in a two-point formula, takes the
Chern-Simons 3-form of the vector potentials derived from the Bloch
wave functions.  Using the relation between polarization and charge
density, we demonstrate our formula by studying charge
fractionalization in a two-dimensional dimer model recently proposed.
\end{abstract}
\maketitle

Electric polarization is a fundamental quantity in condensed matter
physics, essential to any proper description of dielectric phenomena
of matter.  Theoretically, it is well established that only the change
in polarization has physical meaning and it can be quantified by
using the Berry phase of the electronic wave
functions~\cite{king-smith1993,resta1994,ortiz1994}.
In practice, the Berry-phase formula is usually expressed in terms of
the Bloch orbitals.  It has been very successful in first-principles
studies of dielectric properties of oxides and other insulating
materials.

While the existing formulation is adequate in periodic insulators, a
theory of polarization for inhomogeneous crystals would find numerous
important applications; for example, in a class of recently discovered
multiferroics, the appearance of electric polarization is always
accompanied by long-wavelength magnetic
structures~\cite{kimura2003,hur2004,lawes2005}.  A number of
phenomenological and microscopic theories have been proposed to
understand this magnetically induced
polarization~\cite{lawes2005,mostovoy2006,
katsura2005,sergienko2006,hu2007}; however, quantitative studies of
this type of problem still remain in a primitive state.  The
fundamental difficulty lies in the fact that the inhomogeneous
ordering breaks the translational symmetry of the crystal so that
Bloch's theorem does not apply.

In this Letter we present a general framework to calculate electric
polarization in crystals with inhomogeneous ordering.  Our theory is
based on the elementary relation between the change in polarization
and integrated bulk current~\cite{resta1994,ortiz1994}.  The latter
can be evaluated using the semiclassical formalism of Bloch electron
dynamics~\cite{sundaram1999}.  We find that, in addition to the
contribution previously obtained for uniform
systems~\cite{king-smith1993}, the polarization contains an extra
contribution proportional to the gradient of the order parameter.
This extra contribution is expressed using the second Chern form of
the Berry curvatures derived from the \emph{local} Bloch functions.
It can also be recast into a two-point formula, which depends only on
the initial and final states, up to an uncertainty quantum after
spatial averaging.  We identify this quantum as the second Chern
number in appropriate units.  In addition, several general conditions
for the inhomogeneity-induced polarization to be nonzero are also
derived.

To demonstrate our theory, we apply our formula to study the problem
of charge fractionalization in a two-dimensional dimer model recently
proposed~\cite{seradjeh2007,chamon2007}.  We show that in this model
fractional charge appears as a result of the ferroelectric domain
walls.  By using the relation between polarization and charge density,
we calculate the total charge carried by a vortex in the dimerization
pattern and compare it to previous
results~\cite{seradjeh2007,chamon2007}.  Our approach has the
advantage that it can be easily incorporated in a band calculation,
while previously one relied on spectral analysis of the Dirac
Hamiltonian performed in the continuum
limit~\cite{seradjeh2007,chamon2007}.

\textit{General formulation}.---Suppose we have an insulating crystal
with an order parameter $\bm m(\bm r)$ that varies slowly in space.
We assume that, at least on the mean-field level, $\bm m(\bm r)$ can
be treated as an external field that couples to an operator in the
Hamiltonian $\cal H$.  Thus, we can formally write $\cal H[\bm m(\bm
r)]$.  As was emphasized in previous
work~\cite{king-smith1993,resta1994,ortiz1994}, only the change in
polarization $\bm P$ between two different states has meaning, and it
is given by~\cite{P-def}
\begin{equation}  \label{start}
\bm P = \int_0^T dt \, \bm j(\bm r, t) \;,
\end{equation}
where $\bm j(\bm r, t)$ is the bulk current density as the system
adiabatically evolves from the initial state $(t = 0)$ to the final
state $(t = T)$.  In other words, we assume that the two states are
connected through a continuous transformation of the Hamiltonian $\cal
H[\bm m(\bm r);\lambda]$ parameterized by a scalar $\lambda$ with
$\lambda(0) = 0$ and $\lambda(T) = 1$.

In order to find the current density $\bm j(\bm r, t)$, we adopt the
formalism of semiclassical dynamics of Bloch
electrons~\cite{sundaram1999}, which is a powerful tool to investigate
the influence of slowly varying perturbations on electron dynamics.
Within this approach, each electron is described by a narrow wave packet
localized around $\bm r_c$ and $\bm k_c$ in the phase space.
If $\bm m(\bm r)$ varies smoothly compared to the width of the wave
packet, it is sufficient to study a family of \emph{local}
Hamiltonians $\cal H_c[\bm m(\bm r_c); \lambda]$ which assumes a fixed
value of the order parameter $\bm m(\bm r_c)$ in the vicinity of $\bm
r_c$.  Since $\cal H_c[\bm m(\bm r_c); \lambda]$ maintains the
periodicity of the unperturbed crystal, its eigenstates have the Bloch
form: $\ket{\psi_n(\bm k, \bm r_c; \lambda)} = e^{i\bm k \cdot \bm r}
\ket{u_n(\bm k, \bm r_c; \lambda)}$, where $\ket{u_n(\bm k, \bm r_c;
\lambda)}$ is the cell-periodic part of the Bloch functions.  Note
that the $\bm r_c$-dependence of $\ket{u_n(\bm k, \bm r_c; \lambda)}$
enters through $\bm m(\bm r_c)$.  We can then expand the wave packet
using these local Bloch functions.  For simplicity, in the following
derivation we shall confine ourselves to the case of non-degenerate
bands and hence omit the band index $n$.

It has been previously shown that the wave packet center satisfies the
following equations of motion (hereafter the subscript $c$ on $\bm
k_c$ and $\bm r_c$ is dropped)~\cite{sundaram1999}
\begin{subequations} \label{EOM}
\begin{align}
\dot{r}_\alpha &= \nabla^k_\alpha \eps - \Omega^{kr}_{\alpha\beta}
\dot{r}_\beta - \Omega^{kk}_{\alpha\beta} \dot{k}_\beta -
\dot{\lambda} \Omega^{k\lambda}_\alpha \;, \\ 
\dot{k}_\alpha &= -\nabla^r_\alpha \eps + \Omega^{rr}_{\alpha\beta}
\dot{r}_\beta + \Omega^{rk}_{\alpha\beta} \dot{k}_\beta +
\dot{\lambda} \Omega^{r\lambda}_\alpha \;,
\end{align}
\end{subequations}
where $\eps$ is the electron energy and we have introduced the
notation $\nabla^k_\alpha = \partial/\partial k_\alpha$ and
$\nabla^r_\alpha = \partial/\partial r_\alpha$.  Summation over
repeated indices is implied throughout our derivation.  Here,
$\bm{\Omega}$ is the Berry curvature obtained from the vector
potential $\bm{\cal A}$ derived from $\ket{u(\bm k, \bm r, \lambda)}$.
For example,
\begin{gather}
\cal A^k_\alpha = \bracket{u|i\nabla^k_\alpha|u} \;, \quad
\cal A^r_\alpha = \bracket{u|i\nabla^r_\alpha|u} \;, \\
\Omega^{kr}_{\alpha\beta} = \nabla^k_\alpha \cal A^r_\beta 
- \nabla^r_\beta \cal A^k_\alpha \;. \label{curvature}
\end{gather}
Other Berry curvatures are similarly defined.  It is noteworthy that
although the vector potential $\bm{\cal A}$ depends on the phase
choice of the wave function $\ket{u(\bm k, \bm r, \lambda)}$, the
Berry curvature $\bm{\Omega}$ is a well-defined gauge-invariant
quantity in the parameter space $(\bm k, \bm r, \lambda)$.

We now turn to the derivation of $\bm P$ using Eq.~\eqref{start}.  The
\emph{electronic} contribution to polarization is given by
\begin{equation} \label{P}
\bm P = -e \int_\text{BZ} d\bm k \int_0^T dt\,D(\bm k, \bm r) \dot{\bm r} \;,
\end{equation}
where $-e$ is the electron charge, and $D(\bm r, \bm k)$ is the
electron density of states, which is modified from its usual value of
$1/(2\pi)^d$ in the presence of the Berry curvature, $D(\bm k, \bm r)
= (1 + \Omega^{kr}_{\alpha\alpha})/(2\pi)^d$~\cite{xiao2005}.

We can solve $\dot{r}_\alpha$ from Eq.~\eqref{EOM} then insert it into
Eq.~\eqref{P}.  Collecting terms proportional to $\dot{\lambda}$ and
keeping those up to first order in the gradient, we
obtain~\cite{monopole}
\begin{equation}
\bm P = \bm P^{(0)} + \bm P^{(1)} \;,
\end{equation}
where $\bm P^{(0)}$ is the zeroth order contribution
\begin{equation}  \label{P0}
P^{(0)}_\alpha = e \intk \int_0^1 d\lambda\,
\Omega^{k\lambda}_\alpha \;,
\end{equation}
and $\bm P^{(1)}$ is the first order contribution
\begin{equation} \label{P1} \begin{split}
P^{(1)}_\alpha &= e \intk \int_0^1 d\lambda \\
&\qquad \Bigl(\Omega^{kr}_{\beta\beta} \Omega^{k\lambda}_\alpha
- \Omega^{kr}_{\alpha\beta} \Omega^{k\lambda}_\beta
+ \Omega^{kk}_{\alpha\beta} \Omega^{r\lambda}_\beta \Bigr) \;.
\end{split} \end{equation}
These are the central results of this work.  We note that $\bm
P^{(0)}$ has been obtained by King-Smith and Vanderbilt for uniform
systems~\cite{king-smith1993}, whereas $\bm P^{(1)}$, being
proportional to the gradient of $\bm m(\bm r)$, only exists in
inhomogeneous crystals.

Two remarks are in order: firstly, although in the above derivation we
have assumed an inhomogeneous order parameter, it is obvious that our
theory is also applicable when the system is subject to a perturbation
of a spatially-varying external field; secondly, we have only
considered the electronic contribution to $\bm P$ here.  When
comparing with experiment, one should also include the ionic
contribution, which is relatively easy to calculate because of its
classical nature.

\textit{Two-point formula}.---We first show that $\bm P^{(1)}$ has the
desired property that it depends only on the initial and final states.
The gauge-invariance of Eq.~\eqref{P1} allows us to evaluate it with
any gauge choice.  In order to carry out the integration over
$\lambda$, we choose the path-independent gauge by requiring that the
phase difference between $\ket{u(\bm k, \bm r, \lambda)}$ and
$\ket{u(\bm k + \bm G, \bm r, \lambda)}$ does not depend on $\lambda$,
where $\bm G$ is a reciprocal lattice vector~\cite{ortiz1994}.
Under this gauge, Eq.~\eqref{P1} can be recast as~\cite{CS}
\begin{equation} \label{two-point}
P^{(1)}_\alpha = e \intk
\Bigl(\cal A^k_\alpha\nabla^r_\beta \cal A^k_\beta
+ \cal A^k_\beta \nabla^k_\alpha \cal A^r_\beta 
+ \cal A^r_\beta \nabla^k_\beta \cal A^k_\alpha \Bigr)  \Big|_0^1\;.
\end{equation}
We recognize that the integrand in the above equation is nothing but
the Chern-Simons 3-form.

\begin{table}
\caption{\label{comp}Comparison between $\bm P^{(0)}$ and $\bm
  P^{(1)}$}
\begin{ruledtabular}
\begin{tabular}{c|cc}
 & Two-point formula & Uncertain quantum \\
\hline
$\bm P^{(0)}$ & Chern-Simons 1-form & First Chern number \\
$\bm P^{(1)}$ & Chern-Simons 3-form & Second Chern number 
\end{tabular}
\end{ruledtabular}
\end{table}

However, we have paid a price for performing the
$\lambda$-integration; namely, the spatially averaged polarization
$\bracket{P^{(1)}_\alpha} = (1/V)\int d\bm r P^{(1)}_\alpha$ resulting
from this two-point formula~\eqref{two-point} can only be determined
modulo a quantum.

To find the size of the quantum, we consider a cyclic change in
$\lambda$.  Let us now assume that the order parameter $\bm m(\bm r)$
is periodic in $\bm r$.  The integral in Eq.~\eqref{P1} (after a
spatial integration) over a closed manifold spanned by $(k_\alpha,
k_\beta, r_\beta, \lambda)$ is an integer called the second Chern
number~\cite{avron1988}.  Since Eq.~\eqref{two-point} does not track
the evolution of $\lambda$, there is no information of how many cycles
$\lambda$ has gone through.  This is the reason why
$\bracket{P^{(1)}_\alpha}$ using Eq.~\eqref{two-point} can only be
determined modulo a quantum.  Assuming $\bm m(\bm r)$ depends on $y$,
we obtain the quantum for $P^{(1)}_x$ in a three-dimensional system:
\begin{equation}
\Delta\bracket{P^{(1)}_x} = \frac{e}{l_y a_z} \;,
\end{equation}
where $l_y$ is the period of $\bm m(y)$ and $a_z$ is the lattice
constant along $\hat{\bm z}$.

Similarly, the zeroth order contribution $\bm P^{(0)}$ can also be
cast into a two-point formula and the uncertain quantum is given by
$e/(a_ya_z)$~\cite{king-smith1993}.  First-principles calculations
show that in real materials $P^{(0)}$ is usually smaller than this
quantum.  Hence the ratio between $P^{(0)}$ and $P^{(1)}$ is roughly
on the order of $l_y/a_y$.  The similarities between $\bm P^{(0)}$ and
$\bm P^{(1)}$ are summarized in Table~\ref{comp}.

\textit{Minimal conditions for a finite $\bm P^{(1)}$}.---We now
evaluate Eq.~\eqref{P1} using a particular path of $\lambda$.  We
write $\cal H[\bm m(\bm r); \lambda] = \cal H[\lambda\bm m(\bm r)]$ so
that $\lambda$ acts like a ``switch'' of the order $\bm m(\bm r)$,
i.e., when $\lambda = 0$ the system is orderless and when $\lambda =
1$ the order is fully developed.  Using the relation $\nabla^r_\alpha
= \nabla^r_\alpha m_\mu \nabla^m_\mu$ and $\nabla^\lambda =
(m_\mu/\lambda) \nabla^m_\mu$, we can recast Eq.~\eqref{P1} as
\begin{equation} \label{final} \begin{split}
P^{(1)}_\alpha &= e m_\mu\nabla^r_\beta m_\nu \intk \int_0^1
\frac{d\lambda}{\lambda} \\
&\qquad \Bigl(\Omega^{km}_{\alpha\mu}\Omega^{km}_{\beta\nu} 
- \Omega^{km}_{\alpha\nu} \Omega^{km}_{\beta\mu} 
+ \Omega^{kk}_{\alpha\beta} \Omega^{mm}_{\mu\nu} \Bigr) \;.
\end{split} \end{equation}
As we shall see below, this equation is very useful in assessing the
general properties of $\bm P^{(1)}$.

Beside having the crystal be inhomogeneous, there are three general
conditions for $\bm P^{(1)}$ to be nonzero according to
Eq.~\eqref{final}: (i) the system must be two-dimensional or higher;
(ii) the order parameter $\bm m(\bm r)$ must have two or more
components; and (iii) the wave function must depend on four or more
\emph{independent} parameters.  These conditions can be obtained by
realizing that the integrand in Eq.~\eqref{final} is actually the
second Chern 4-form $\Omega \wedge \Omega$ given in its local
expression with respect to the coordinates $(k_\alpha, k_\beta, m_\mu,
m_\nu)$.  It is antisymmetric in $k_\alpha$ and $k_\beta$, and in
$m_\mu$ and $m_\nu$, hence condition (i) and (ii).  Condition (iii)
follows from the fact that all 4-forms vanish identically in three or
less dimensions.  Based on condition (iii) we can further deduce that
$\dim(\cal H) > 2$.  If $\dim(\cal H) = 2$, $\cal H$ has four
components.  However, since shifting and scaling energy has no effect
on wave functions, the wave function can depend on only two
\emph{independent} parameters (for example, the spherical coordinates
on a 2-sphere $S^2$) and $\bm P^{(1)}$ vanishes in this case.  This
set of conditions puts powerful constraints on possible microscopic
models that display finite $\bm P^{(1)}$.  Conditions (i) and (iii)
can also be obtained directly from Eq.~\eqref{P1}.

Let us consider a two-dimensional ``minimal'' model and assume that
both the space of $\bm m(\bm r)$ and coordinate space are
two-dimensional.  Because of its antisymmetric properties, we can
write the integrand of Eq.~\eqref{final} as $\epsilon_{\alpha\beta}
\epsilon_{\mu\nu} \chi$.  Then Eq.~\eqref{final} takes the following
form
\begin{equation} \label{2d}
\bm P^{(1)} = e\chi[(\bm\nabla \cdot \bm m) \bm m - (\bm m \cdot
  \bm\nabla) \bm m] \;,
\end{equation}
Here $\chi$, as a function of $\bm m(\bm r)$, can be spatial
dependent.  Interestingly, if we identify $\bm m(\bm r)$ with the
magnetization order parameter $\bm M(\bm r)$, the above result is
consistent with the Landau-Ginzburg theory of polarization induced by
spiral magnetic ordering~\cite{mostovoy2006}.  However, our
result~\eqref{2d} is a direct consequence of the minimal
dimensionality and we did not invoke any symmetry analysis.  For
higher dimensions, one will have to carry out a careful symmetry
analysis of the magnetic groups of the crystal~\cite{harris2007}.

\textit{Degenerate bands}.---So far, our derivation is for
non-degenerate bands. The generalization to degenerate bands is
straightforward~\cite{culcer2005,shindou2005}.  The vector potential
and Berry curvature become matrix-valued and are defined by
\begin{gather}
(\cal A_a)_{mn} = \bracket{u_m|i\nabla_a|u_n} \;, \\
\Omega_{ab} = \nabla_a \cal A_b - \nabla_b \cal A_a
- i[\cal A_a, \cal A_b] \;,
\end{gather}
where $a, b \in (\bm k, \bm r, \lambda)$ and $\ket{u_m}$ and
$\ket{u_n}$ are degenerate bands.  We then need to take the trace of
Eqs.~\eqref{P0} and \eqref{P1} for the zeroth and first order
contributions to $\bm P$.  The two-point formula in Eq.~\eqref{two-point}
also takes the non-Abelian Chern-Simons form.

\begin{figure}
\includegraphics[width=8.5cm]{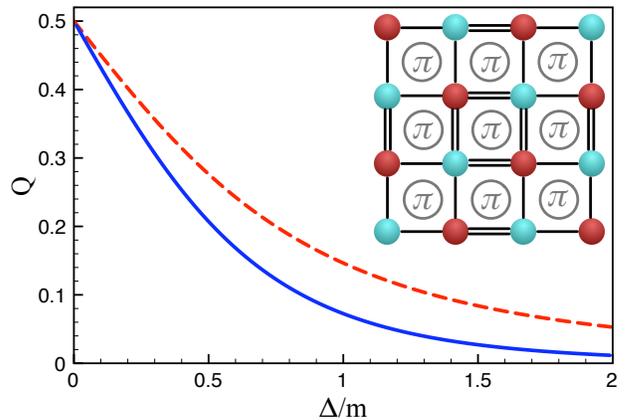}
\caption{\label{fig:frac}(color online). The total charge carried by
  the ferroelectric vortex domain wall $m(\bm r) e^{i\theta} = m_x + i
  m_y$ as a function of $\Delta/m$, where $\Delta$ is the staggered
  sublattice potential, and $m$ is the dimerization order parameter.
  As $\Delta$ increases, the difference between the result from the
  continuum limit (red dashed line) and that based on the band
  calculation (blue solid line) becomes significant.  The insert shows
  the two-dimensional dimerized square lattice with $\pi$-flux per
  plaquette in the absence of the vortex.  The hopping amplitude is
  $t(1\pm m_{x,y})$ along the $x,y$-direction.}
\end{figure}

\textit{Fractional charge}.---To demonstrate our theory, we consider
the problem of charge fractionalization in a recently proposed
two-dimensional dimer model~\cite{seradjeh2007,chamon2007}, shown
schematically in the inset of Fig.~\ref{fig:frac}.  Introducing
$\gamma_i = \sigma_i \otimes \sigma_z$, $\gamma_4 = \openone \otimes
\sigma_x$ and $\gamma_5 = \openone \otimes \sigma_y$, we can write the
Hamiltonian as $\cal H = h_\alpha \gamma_a$, where
\begin{equation} \label{ham}
\bm h = t(\cos k_x, \cos k_y, \Delta, m_x \sin k_x,  m_y \sin k_y) \;,
\end{equation}
$t\Delta$ is the staggered sublattice potential, $t(1 \pm m_x)$ and
$t(1 \pm m_y)$ are the dimerized hopping amplitudes along $x$ and $y$
direction.  We choose the Landau gauge so that the effect of the $\pi$
flux is represented by alternating signs of the hopping amplitudes
along adjacent rows.  It turns out that this model is a minimal one
satisfying all our three conditions: (i) it is two-dimensional;
(ii) the order parameter $\bm m= (m_x, m_y)$ has two components; and
(iii) $\bm h$ (after scaling) can be mapped onto a unit sphere $S^4$
with four independent spherical angles.

It can be verified that the energy spectrum of this Hamiltonian
consists of two doubly degenerate levels; therefore, the non-Abelian
formalism is necessary.  The Berry curvature has SU(2)
symmetry~\cite{avron1988,shankar1994,demler1999}; hence, $\bm P^{(0)}$
always vanishes since the non-Abelian version of Eq.~\eqref{P0} has
vanishing trace.  Thus, we will only consider $\bm P^{(1)}$ in what
follows.

Suppose there is a vortex in the dimerization pattern: namely, $m_x +
im_y = m(r) e^{in\theta}$.  According to Eq.~\eqref{2d} together with
the fact that $\rho(\bm r) = -\bm\nabla\cdot\bm P$, this ferroelectric
vortex domain wall will carry a polarization charge of $Q = \int d\bm
r \rho(\bm r) = n m^2\int_0^{2\pi} d\theta\,\chi$~\cite{mostovoy2006},
which is in general fractional.

To compare with previous results, we shall first evaluate $\chi$ in the
continuum limit.  Expanding the Hamiltonian around the Dirac point
$(\pi/2, \pi/2)$, we find, according to Eq.~\eqref{final},
\begin{equation}
\chi = \frac{3}{2} n\int \frac{d\bm k}{(2\pi)^2} \int_0^1 d\lambda
\frac{\Delta \lambda}{(k^2 + m^2\lambda^2 + \Delta^2)^{5/2}} \;.
\end{equation}
Since at large $k$ the integrand decays as $k^{-5}$, we can extend the
integration range of $\bm k$ to infinity and obtain
\begin{equation}
\chi = \frac{n}{4\pi m^2}(1 - \frac{\Delta}{\sqrt{\Delta^2 + m^2}}) \;,
\end{equation}
and the total charge carried by the vortex is given by
\begin{equation}
Q = n\frac{e}{2}(1 - \frac{\Delta}{\sqrt{\Delta^2 + m^2}}) \;,
\end{equation}
where $n$ is the winding number.  This result agrees with the spectral
analysis of the Dirac Hamiltonian~\cite{seradjeh2007,chamon2007}.  

The above derivation provides a simple picture of charge
fractionalization in this type of system: it is a direct consequence
of the ferroelectric domain wall, and the breaking of the sublattice
symmetry ($\Delta$) allows it to be irrational.  A detailed report
including both 1D and 2D cases will be reported elsewhere.  We also
calculate the total charge based on a band calculation using
Eq.~\eqref{ham}, shown in Fig.~\ref{fig:frac}.  As $\Delta$ increases,
the deviation between the band calculation and continuum limit becomes
significant.

In summary, we have developed a general theory of polarization induced by
inhomogeneity in crystals.  Our result lays the foundation for
quantitative studies of this type of problem.  In connection to
multiferroics, the minimal conditions for a finite $\bm P^{(1)}$ point
to general directions to aid in the search for microscopic models.  In
addition, we have illustrated our theory by showing that the fractional
charge in certain models can be understood as the polarization charge
accompanying ferroelectric domain walls.

DX thanks D.~Culcer for useful discussions.  DX was supported by the
NSF (DMR-0404252/0606485), JRS by the NSF of China (No.~10604063), DPC
by DARPA (No.~MDA0620110041), and QN by the Welch Foundation, DOE
(DE-FG03-02ER45958), and the NSF of China (No.~10740420252).


\begin{thebibliography}{24}
\bibitem{P-def}
  Strictly speaking, there is still an ambiguity in the definition of
  $\bm P$ given here because $\bm j = \partial \bm P/\partial t +
  \bm\nabla \times \bm M$ also contains a contribution from the
  magnetization current.  As a result, $\bm P$ is only defined up to a
  divergence-free field.  One can of course fix the gauge by imposing
  that $\bm P$ only have a longitudinal component.  However, this is
  not a critical issue because the ambiguity can be removed by spatial
  averaging.
\bibitem{monopole}
  The integral of terms that do not contain $\dot{\lambda}$ is given
  by $-e \int_\text{BZ} d\bm k \int_0^T dt [\nabla^k_\alpha \eps +
  (\Omega^{kr}_{\beta\beta} \nabla^k_\alpha \eps -
  \Omega^{kr}_{\alpha\beta} \nabla^k_\beta \eps +
  \Omega^{kk}_{\alpha\beta} \nabla^r_\beta \eps)]$.  The integral of
  $\nabla^k_\alpha \eps$ over the entire Brillouin zone obviously
  vanishes.  After integration by parts and making use of the Bianchi
  identity $\nabla^k_\alpha \Omega^{rk}_{\beta\beta} + \nabla^k_\beta
  \Omega^{kr}_{\alpha\beta} + \nabla^r_\beta \Omega^{kk}_{\beta\alpha}
  = 0$, one can show that the integral of the last three terms
  contributes a divergence-free part, which can be discarded.
\bibitem{CS} 
  The formal derivation of Eq.~\eqref{two-point} is lengthy
  and will be reported elsewhere.  In the simplest case where
  $\bm{\cal A}$ is well-defined everywhere in $(\bm k, \bm r,
  \lambda)$, one can prove Eq.~\eqref{two-point} by using Stokes'
  theorem $\int_{\cal M} \Omega \wedge \Omega = \int_{\cal M} d(\cal A
  \wedge d\cal A) = \int_{\partial\cal M} \cal A \wedge d\cal A$.
\end{thebibliography}
\end{document}